\documentclass[twocolumn,amsmath,amssymb,superscriptaddress,notitlepage]{revtex4-1}
\usepackage{graphics}
\usepackage{bm}
\usepackage{dcolumn}
\usepackage{natbib}
\usepackage{epstopdf}
\usepackage{float}
\usepackage{graphicx}
\usepackage{epsfig}
\usepackage[pdfstartview=FitH]{hyperref}
\usepackage{color}
\usepackage{appendix}
\usepackage{ulem}

\newcommand{\qq}{{\bf q}}
\newcommand{\rr}{{\bf r}}

\begin{document}
\title{Integer quantum Hall effect of two-component hardcore bosons in a topological triangular lattice}
\author{Tian-Sheng Zeng}
\affiliation{Department of Physics, College of Physical Science and Technology, Xiamen University, Xiamen 361005, China}
\date{\today}
\begin{abstract}
We study the many-body ground states of two-component hardcore bosons in topological triangular lattice models. Utilizing exact diagonalization and density-matrix renormalization group calculations, we demonstrate that at commensurate two-thirds filling per lattice site, a two-component bosonic integer quantum Hall (BIQH) effect emerges with the associated $\mathbf{K}=\begin{pmatrix}
0 & 1\\
1 & 0\\
\end{pmatrix}$ matrix under strong intercomponent Hubbard repulsion. The topological nature is further elucidated by (i) a unique ground state degeneracy with a robust spectrum gap, (ii) a quantized topological Chern number matrix $\mathbf{C}=\mathbf{K}^{-1}$, and (iii) two counterpropagating edge branches. Moreover, with increasing nearest-neighbor repulsions, the ground state undergoes a first-order transition from a BIQH liquid to a commensurate solid order.
\end{abstract}
\maketitle

\section{Introduction}

The bosonic integer quantum Hall (BIQH) state, as a bosonic analog of a fermionic topological insulator, is one of the simplest symmetry-protected topological phases with $U(1)\times U(1)$ global symmetry~\cite{Senthil2013}. In the past, a possible integer quantized quantum Hall effect was numerically considered in Harper-Hofstadter model with topological $C=2$ band~\cite{Cooper2009}. Later such numerical studies were extended to strongly interacting two-component bosons in two dimensional magnetic field based on cold atomic neutral systems~\cite{Furukawa2013,Regnault2013,Wu2013,Grass2014}. The fascinating properties of a BIQH phase are that it is characterized by (a) Hall conductivity quantized to an even integer~\cite{Lu2012,Wen2013,Geraedts2013} and (b) two counter propagating chiral modes. Recently, such lattice realizations of a BIQH phase have been also actively explored, including exact diagonalization (ED) calculation of single-component bosons at filling $\nu=1$ in the lowest topological flat-band with Chern number $C=2$ identified by a unique ground state with a quantized $|\sigma_{xy}=2|$ Hall conductance~\cite{Sterdyniak2015,Zeng2016}, density-matrix renormalization group (DMRG) identification of various hardcore bosonic lattices with correlated hoppings through Laughlin's argument of quantized charge pumping related to Hall conductance~\cite{He2015,Chen2019,Zeng2020}, and Monte Carlo study of two interpenetrating square lattices of quantum rotors~\cite{Geraedts2017}.
In Ref.~\cite{Cooper2015}, composite fermion theory was proposed for the symmetry-protected BIQH state of single component bosons in Chern bands with $C=2$.

However so far examples of the BIQH phase for two-component bosons in Chern bands with $C=1$ are lacking, and the identification of such a symmetry-protected topological insulator in bosonic systems would expand the taxonomy of topological phases of matter, which is the aim of our current paper. Actually for single-component bosons in topological flat bands, previous numerical investigations reveal an analog of quantum Hall effects in Chern bands with $C=1$ to those in continuum Landau levels~\cite{Sun2011,Neupert2011,Sheng2011,Tang2011,Wang2011,Regnault2011}, and a series of color-entangled Abelian topological states in Chern bands with $C>1$ at various filling fractions~\cite{LBFL2012,Wang2012r,Yang2012,Sterdyniak2013,Wang2013}. Nevertheless, the topological order of multicomponent bosons in topological flat bands reveals a fertile topic to be uncovered with many possible intriguing topological phases~\cite{Wen2016}, aside from the theoretical interest. In a series of works, quantum Hall effects of multicomponent bosons including Halperin ($mmn$) states~\cite{Zeng2017,Zeng2020b} and their multicomponent generalizations (including Bose-Fermi mixtures and non-Abelian spin-singlet clusters)~\cite{Zeng2018,Zeng2019,Zeng2021,Zeng2022} in topological flat bands with $C=1$ have been numerically demonstrated through ED and DMRG calculations of both intracomponent and intercomponent Hall transport responses. The versatile experimental ability in the design and control of different iconic topological models in cold atoms, such as Haldane-honeycomb~\cite{Jotzu2014} and Harper-Hofstadter models~\cite{Aidelsburger2013,Miyake2013,Mancini2015,Stuhl2015}, has been exemplified as a valuable platform for studying such topological phases in Chern bands for bosons in cold atoms systems.

In this paper, we theoretically address the open issues regarding the emergence of a robust BIQH state at filling factor $\nu=2$ in certain topological lattices, and then further validate its topological properties. This paper is organized as follows. In Sec.~\ref{model}, we introduce the interacting two-component hardcore bosonic Hamiltonian in a three-band triangular-lattice model with the lowest Chern band of Chern number $C=1$ under onsite and nearest-neighboring Hubbard repulsions, and give a description of our numerical methods. In Sec.~\ref{biqh}, we explore the many-body ground state at strong onsite Hubbard repulsion and present a detailed proof of BIQH state at filling $\nu=2$ by ED and DMRG calculations of its topological information. We will explore its ground state degeneracy, Chern number matrix, drag charge pumping and chiral edge modes, according to the $\mathbf{K}$ matrix classification. Also we examine the possible competing solid order phase arising from nearest-neighboring repulsion, and discuss its transition into BIQH state by varying interactions. Finally, in Sec.~\ref{summary}, we summarize our results and discuss the primary interplay between interaction and band topology.

\section{Model and Method}\label{model}

\begin{figure}[t]
  \includegraphics[height=1.7in,width=3.3in]{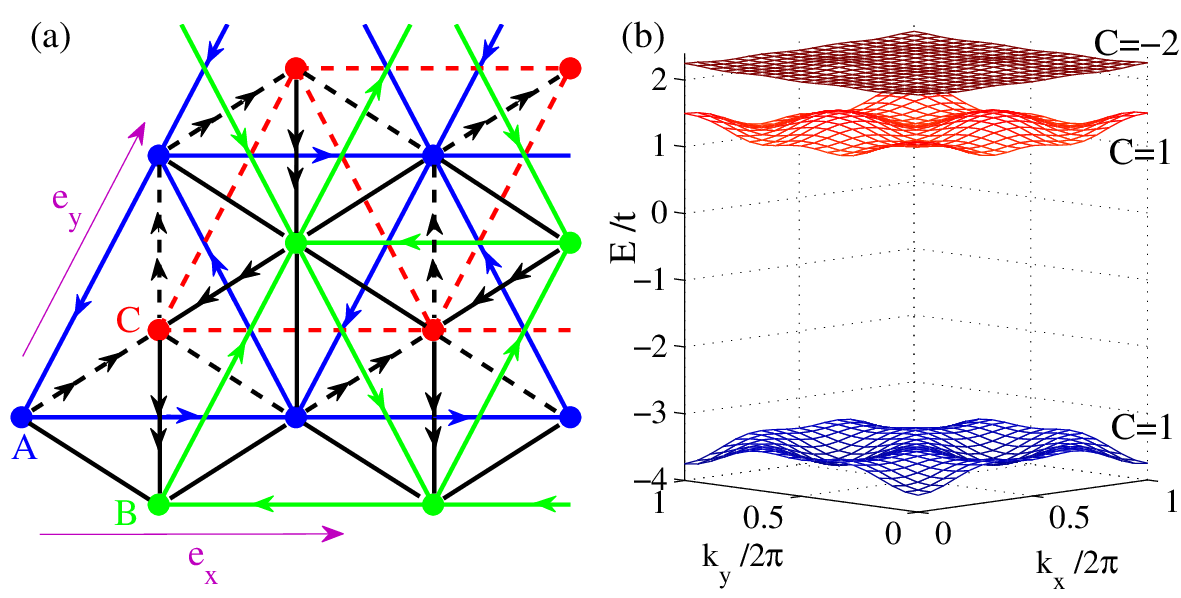}
  \caption{\label{lattice}(Color online) (a) Schematic plot of the topological triangular lattice model in Eq.~\ref{tri}. The double-arrow link shows the hopping direction carrying chiral flux $\phi_{\rr,\rr'}=2\phi$ in the nearest-neighbor hopping, while the single-arrow link shows the hopping direction carrying chiral flux $\phi_{\rr,\rr'}=\phi$ in the next-nearest-neighbor hopping. Three sublattices $A,B,C$ are labeled by blue, green and red solid circles respectively. The hopping amplitudes are $t_1,t_2$ ($-t_1,-t_2$) along the solid (dotted) lines. The magenta $e_{x},e_{y}$ indicate the real-space lattice translational vectors. (b) Single-particle energy spectrum in the Brillouin zone with the Chern number of each band.}
\end{figure}

Here, we will consider the following Hamiltonian of interacting two-component hardcore bosons coupled with each other via onsite and extended Hubbard interactions in a three-band topological triangular lattice model,
\begin{align}
  H&=\!\sum_{\sigma}\!\Big[\pm t_1\!\!\sum_{\langle\rr,\rr'\rangle}\! e^{i\phi_{\rr'\rr}}b_{\rr',\sigma}^{\dag}b_{\rr,\sigma}+H.c.\nonumber\\
  &\pm t_2\sum_{\langle\langle\rr,\rr'\rangle\rangle}\!\!\! b_{\rr',\sigma}^{\dag}b_{\rr,\sigma}e^{i\phi_{\rr'\rr}}+H.c.\Big]\nonumber\\
  &+U\sum_{\rr}n_{\rr,\uparrow}n_{\rr,\downarrow}+V\sum_{\sigma,\sigma'}\!\sum_{\langle\rr,\rr'\rangle}n_{\rr',\sigma'}n_{\rr,\sigma}.\label{tri}
\end{align}
Here $\sigma=\uparrow,\downarrow$ are the pseudospin indices for hardcore bosons (for instance, spinor bosons or a bilayer system), $b_{\rr,\sigma}^{\dag}$ is the particle creation operator of pseudospin $\sigma$ at site $\rr$, $n_{\rr,\sigma}=c_{\rr,\sigma}^{\dag}c_{\rr,\sigma}$ is the particle number of pseudospin $\sigma$ at site $\rr$ ($n_{\rr,\sigma}$ takes the values 0 or 1 due to the hardcore constraint), $\langle\ldots\rangle$ and $\langle\langle\ldots\rangle\rangle$ denote the nearest-neighbor and next-nearest-neighbor pairs of sites. The geometry of topological triangular lattice is depicted in Fig.~\ref{lattice}(a) with three inequivalent sites $A,B,C$ in each unit cell. And the total number of lattice sites $N_s=3\times N_x\times N_y$ with $N_x\times N_y$ unit cells.

However in contrast to Ref.~\cite{Wang2012r}, it is emphasized that here we choose the negative tunnel couplings $t_1=-t,t_2=-t/4,t=1$ with the same chiral flux structure $\phi=\pi/3$, such that the single-particle band structure is reverted with the lowest topological flat band of Chern number $C=1$. The peculiar property is that the lowest Chern band is well-separated from the upper bands with a very large energy gap around $\Delta/t=4.7$ and the mixing effect of the upper bands could be small even for strong interacting particles, as indicated in Fig.~\ref{lattice}(b). Further we take the strength of the onsite Hubbard repulsion $U$, and the strength of the nearest-neighbor repulsion $V$.

\begin{figure}[t]
  \includegraphics[height=1.82in,width=3.3in]{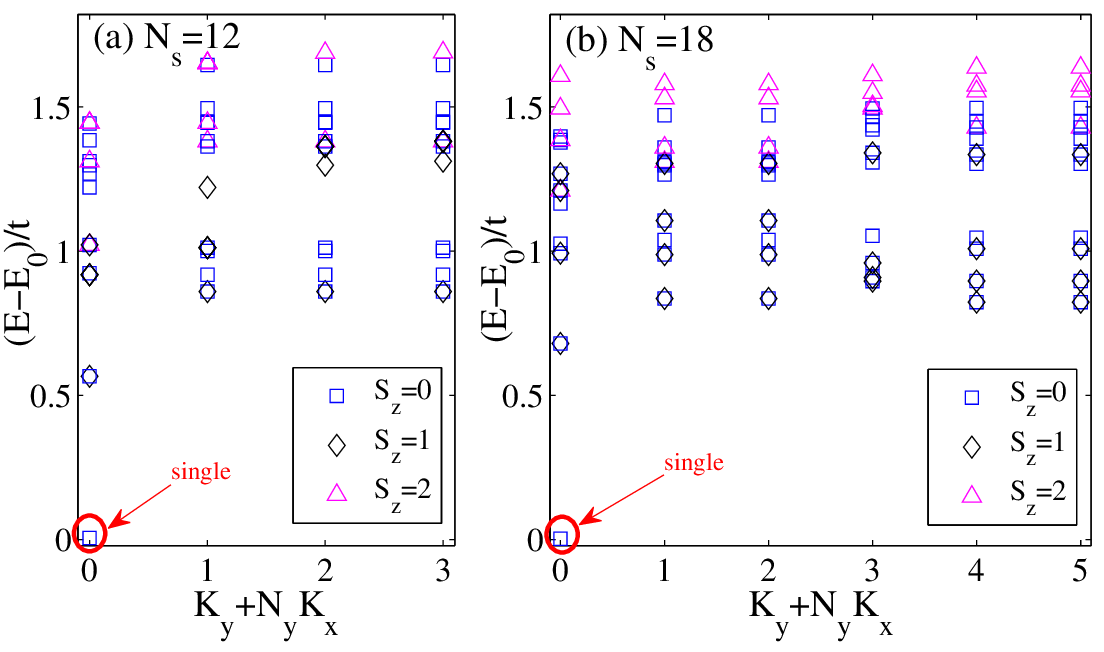}
  \caption{\label{energy} (Color online) Numerical ED results for the low energy spectrum of two-component hardcore bosons at $\nu=2$ with $U=\infty,V=0$ in topological triangular lattice. The system sizes (a) $N_s=3\times2\times2$ and (b) $N_s=3\times2\times3$. }
\end{figure}

In our exact diagonalization (ED) and density-matrix renormalization group (DMRG) simulations, we fix the total particle filling of the lowest Chern band at $\nu=\nu_{\uparrow}+\nu_{\downarrow}=2$ with $\nu_{\uparrow}=N_{\uparrow}/(N_xN_y),\nu_{\downarrow}=N_{\downarrow}/(N_xN_y)$, where $N_{\sigma}$ is the global particle number of pseudospin $\sigma$ with $U(1)$-symmetry. Thus the occupation filling per lattice site is commensurate $\langle n_{\rr}\rangle=\langle \sum_{\sigma}n_{\rr,\sigma}\rangle=(N_{\uparrow}+N_{\downarrow})/N_s=2/3$ in triangular lattice.

In the ED study of small finite periodic lattice systems, the energy states are labeled by the total momentum $K=(K_x,K_y)$ in units of $(2\pi/N_x,2\pi/N_y)$ in the Brillouin zone. While the ED calculations on torus geometry are limited to the system with 18 sites, we exploit infinite DMRG on the cylindrical geometry for larger systems, and keep the maximal bond dimension up to $M=14000$ to obtain accurate results, starting with a random initial state. In the DMRG, we choose the geometry of cylinders with open boundary conditions along the $x$ direction, and periodic boundary conditions along the $y$ direction.

\section{Bosonic Integer Quantum Hall Effect}\label{biqh}

In this section, we set out to systematically present and discuss numerical results for the topological information of the many-body ground state of the model Hamiltonian Eq.~\ref{tri}. For $\nu=2$ filling of topological flat bands, we anticipate bosonic integer quantum Hall effect, while for commensurate filling $2/3$ in triangular lattice, the competing solid order is also expected yet. Without nearest-neighboring interactions $V=0$, below we show that two-component BIQH effect emerges under strong onsite intercomponent repulsion $U\gg t$, whose topological order is classified by the $\mathbf{K}=\begin{pmatrix}
0 & 1\\
1 & 0\\
\end{pmatrix}$ matrix. In the following part we shall elucidate the topological properties of the ground state, including topological degeneracy, topological Chern number matrix, drag charge pumping, and entanglement spectrum, according to the field theory predictions of the characteristic $\mathbf{K}$ matrix.

\subsection{Ground state degeneracy}

\begin{figure}[t]
  \includegraphics[height=1.85in,width=3.4in]{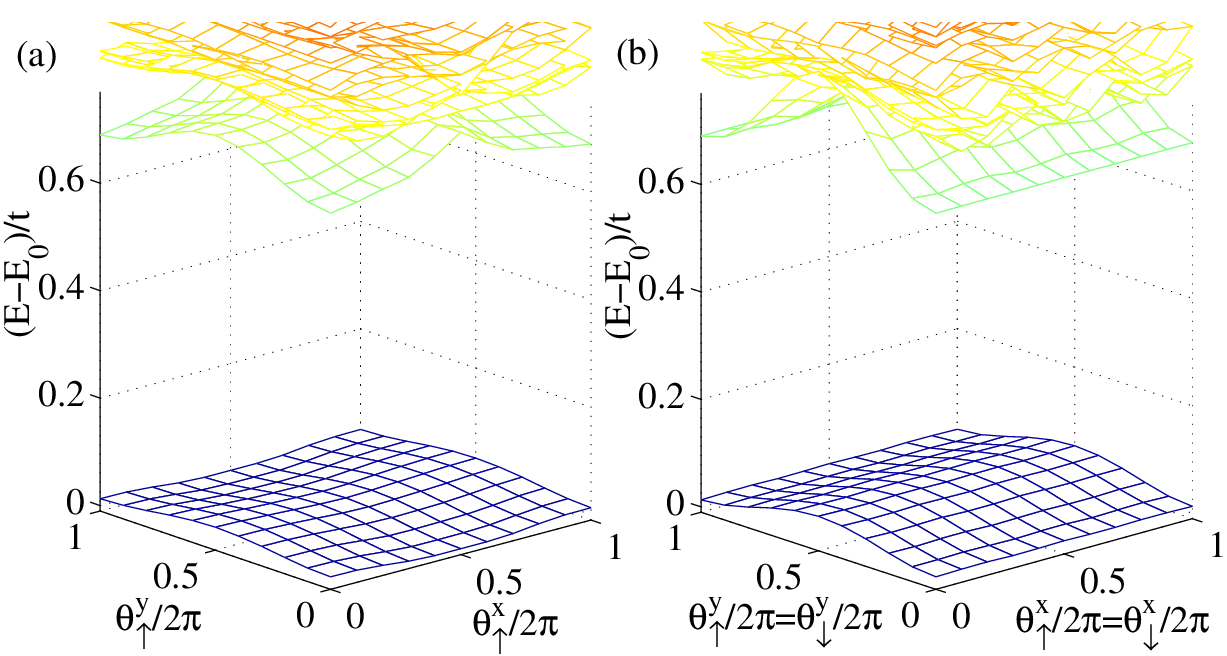}
  \caption{\label{flux} (Color online) Numerical ED results for the low energy spectral flow of two-component hardcore bosons $N_{\uparrow}=6,N_{\downarrow}=6,N_s=18$ with $U=\infty,V=0$ in topological triangular lattice on the two different parameter planes: (a) $(\theta_{\uparrow}^{x},\theta_{\uparrow}^{y})$ with $\theta_{\downarrow}^{x}=\theta_{\downarrow}^{y}=0$ and (b) $(\theta_{\uparrow}^{x}=\theta_{\downarrow}^{x},\theta_{\uparrow}^{y}=\theta_{\downarrow}^{y})$.}
\end{figure}

First, we demonstrate the topological ground state degeneracy on finite periodic lattice using the ED study. As shown in Figs.~\ref{energy}(a) and~\ref{energy}(b) for two-component hardcore bosons in a strongly interacting regime $U\gg t,V=0$ (i.e. the infinite case $U=\infty$ which avoids the double occupancy by any component and has the merit of reducing the size of the Hilbert space in numerical calculations), our extensive numerical calculations of the low energy spectrum in different finite periodic lattice systems show that there exists a well-defined single ground state at momentum sector $K=(0,0)$, which is separated from higher energy levels by a large visible gap. We also calculate the density and spin structure factors for the ground state using DMRG on infinite cylinders, and exclude any possible charge or spin density wave orders as the competing ground states, due to the featureless constant value in the correlation functions $\langle n_{\rr,\sigma}n_{\rr',\sigma'}\rangle\simeq\langle n_{\rr,\sigma}\rangle\langle n_{\rr',\sigma'}\rangle\simeq1/9$ for $|\rr'-\rr|>2$.

Meanwhile to demonstrate the topological robustness of the ground state, we use the twisted boundary condition $\psi(\rr_{\sigma}+N_{\alpha})=\psi(\rr_{\sigma})\exp(i\theta_{\sigma}^{\alpha})$ where $\theta_{\sigma}^{\alpha}$ is the twisted angle of pseudospin $\sigma$ in the $\alpha=x,y$ direction, mimicking the particle momentum shift $k_{\alpha}^{\sigma}\rightarrow k_{\alpha}^{\sigma}+\theta_{\sigma}^{\alpha}/N_{\alpha}$. We further plot the evolution of the low energy spectral flow on the parameter plane of two different types of flux quanta $(\theta_{\uparrow}^{x},\theta_{\uparrow}^{y})$ with $\theta_{\uparrow}^{x}=\theta_{\downarrow}^{y}=0$ and $(\theta_{\uparrow}^{x}=\theta_{\downarrow}^{x},\theta_{\uparrow}^{y}=\theta_{\downarrow}^{y})$. As shown in Figs.~\ref{flux}(a) and~\ref{flux}(b), we find that this unique ground state at $K=(0,0)$ evolves without mixing with any excited level, and the system always returns back to itself upon the insertion of one flux quanta when $\theta_{\uparrow}^{\alpha}=2\pi,\theta_{\downarrow}^{\alpha}=0$
and $\theta_{\uparrow}^{\alpha}=\theta_{\downarrow}^{\alpha}=2\pi$. This robust unique degeneracy is consistent with the determinant of the $\mathbf{K}$ matrix.

\subsection{Chern number matrix}

\begin{figure}[t]
  \includegraphics[height=1.82in,width=3.4in]{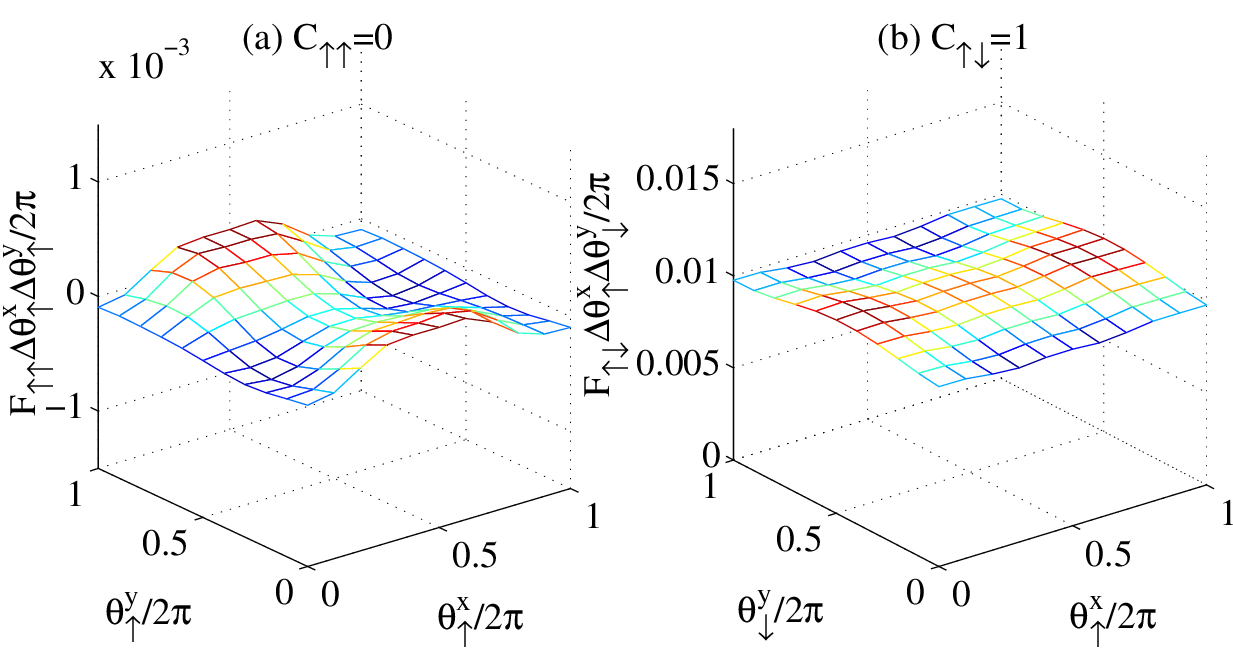}
  \caption{\label{berry} (Color online) Numerical ED results for Berry curvatures $F_{\sigma,\sigma'}^{xy}\Delta\theta_{\sigma}^{x}\Delta\theta_{\sigma'}^{y}/2\pi$ of the unique ground state at $K=(0,0)$ of two-component hardcore bosons $N_{\uparrow}=6,N_{\downarrow}=6,N_s=18$ with $U=\infty,V=0$ in topological triangular lattice under different twisted angles: (a) $(\theta_{\uparrow}^{x},\theta_{\uparrow}^{y})$ and (b) $(\theta_{\uparrow}^{x},\theta_{\downarrow}^{y})$.}
\end{figure}

We numerically divide the twisted angles $(\theta_{\sigma}^{x},\theta_{\sigma'}^{y})$ into $(m+1)\times(m+1)$ coarsely discretized mesh points $(\theta_{\sigma}^{x},\theta_{\sigma'}^{y})=(2k\pi/m,2l\pi/m)$ where $0\leq k,l\leq m$. Then the Berry connection of the wavefunction between two neighboring mesh points is defined as $A_{k,l}^{\pm x}=\langle\psi(k,l)|\psi(k\pm1,l)\rangle$, $A_{k,l}^{\pm y}=\langle\psi(k,l)|\psi(k,l\pm1)\rangle$.
And the Berry curvature on the small Wilson loop plaquette $(k,l)\rightarrow(k+1,l)\rightarrow(k+1,l+1)\rightarrow(k,l+1)\rightarrow(k,l)$ is given by the gauge-invariant expression $F_{\sigma\sigma'}(\theta_{\sigma}^{x},\theta_{\sigma'}^{y})\times4\pi^2/m^2=\mathbf{Im}\ln
\big[A_{k,l}^{x}A_{k+1,l}^{y}A_{k+1,l+1}^{-x}A_{k,l+1}^{-y}\big]$.

From this we can analyze the quantized topological Chern number of a given many-body ground state for interacting systems, in relation to Hall conductance~\cite{Niu1985}. For two-component systems, we adopt the Chern number matrix $\mathbf{C}=\begin{pmatrix}
C_{\uparrow\uparrow} & C_{\uparrow\downarrow} \\
C_{\downarrow\uparrow} & C_{\downarrow\downarrow} \\
\end{pmatrix}$ which is previously introduced for quantum spin Hall effect~\cite{Sheng2003,Sheng2006}. On the parameter plane of two independent twisted angles $\theta_{\sigma}^{x}\subseteq[0,2\pi],\theta_{\sigma'}^{y}\subseteq[0,2\pi]$, we can also define the Chern number of the many-body ground state wavefunction $\psi(\theta_{\sigma}^{x},\theta_{\sigma'}^{y})$ as an integral $C_{\sigma\sigma'}=\int\int d\theta_{\sigma}^{x}d\theta_{\sigma'}^{y}F_{\sigma\sigma'}(\theta_{\sigma}^{x},\theta_{\sigma'}^{y})/2\pi$, where the Berry curvature is given by
\begin{align}
  F_{\sigma\sigma'}(\theta_{\sigma}^{x},\theta_{\sigma'}^{y})=\mathbf{Im}\left(\langle{\frac{\partial\psi}{\partial\theta_{\sigma}^x}}|{\frac{\partial\psi}{\partial\theta_{\sigma'}^y}}\rangle
-\langle{\frac{\partial\psi}{\partial\theta_{\sigma'}^y}}|{\frac{\partial\psi}{\partial\theta_{\sigma}^x}}\rangle\right).\nonumber
\end{align}
In our ED study of finite system sizes for two-component hardcore bosons at $\nu=2$, for the unique ground state at momentum $K=(0,0)$, by numerically calculating the Berry curvatures using $m\times m$ mesh Wilson loop plaquette in the boundary phase space with $m\geq10$, we obtain the topological invariant as a summation over these discretized Berry curvatures. As indicated in Figs.~\ref{berry}(a) and~\ref{berry}(b),
we confirm that for different finite sizes, this unique ground state hosts a vanishing diagonal Chern number $C_{\uparrow\uparrow}=0$ with a vanishingly small Berry curvature $|F_{\uparrow\uparrow}^{xy}\Delta\theta_{\uparrow}^{x}\Delta\theta_{\uparrow}^{y}/2\pi|\ll1$, and an integer quantized off-diagonal Chern number $C_{\uparrow\downarrow}=1$ with a finite smooth Berry curvature $F_{\uparrow\downarrow}^{xy}\Delta\theta_{\uparrow}^{x}\Delta\theta_{\downarrow}^{y}/2\pi$. Therefore, we establish that the ground state hosts a well-defined Chern number matrix, which just equals to the inverse of the $\mathbf{K}=\begin{pmatrix}
0 & 1 \\
1 & 0 \\
\end{pmatrix}$ matrix,
\begin{align}
  \mathbf{C}=\begin{pmatrix}
C_{\uparrow\uparrow} & C_{\uparrow\downarrow} \\
C_{\downarrow\uparrow} & C_{\downarrow\downarrow} \\
\end{pmatrix}=\begin{pmatrix}
0 & 1 \\
1 & 0 \\
\end{pmatrix}=\mathbf{K}^{-1}.\label{chern}
\end{align}

\begin{figure}[t]
  \includegraphics[height=1.78in,width=3.3in]{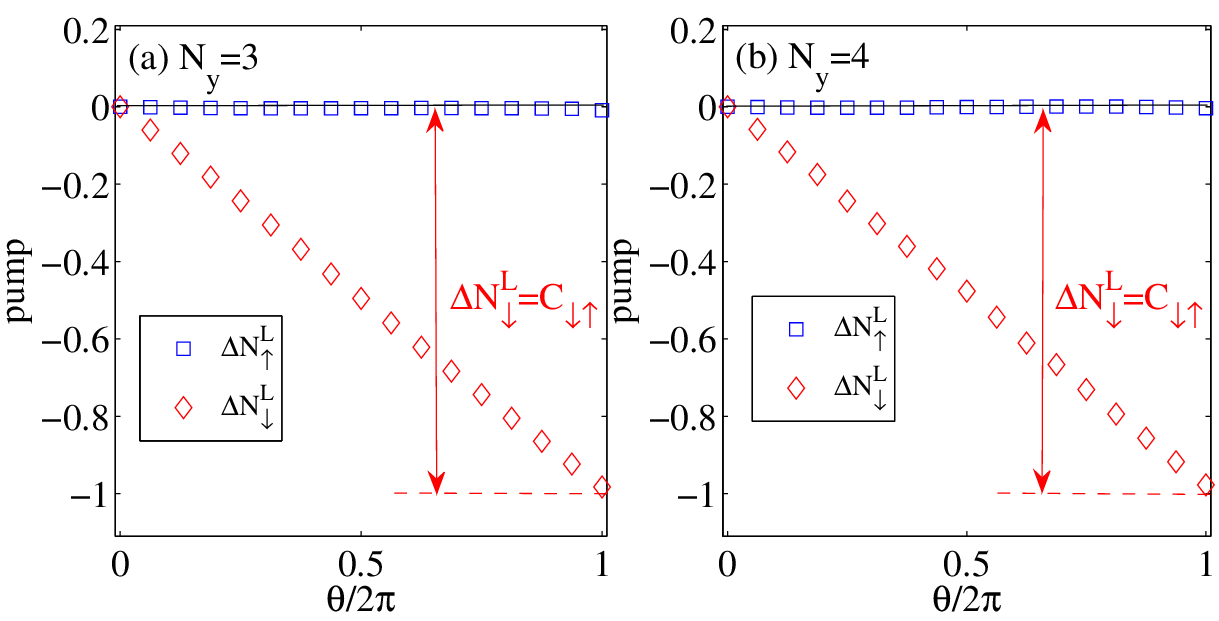}
  \caption{\label{pump} (Color online) The charge transfer in the $x$ direction for two-component hardcore bosons at $\nu_{\uparrow}=\nu_{\downarrow}=1$ with $U=\infty,V=0$ on the infinite cylinder of topological triangular lattice at two-third lattice filling under the insertion of flux
  quantum $\theta_{\uparrow}^{y}=\theta,\theta_{\downarrow}^{y}=0$ of pseudospin $\uparrow$ along the $y$ direction for different cylinder widths: (a) $N_y=3$ and (b) $N_y=4$.}
\end{figure}

\subsection{Drag charge pumping}

Furthermore, as a remarkable feature, Eq.~\ref{chern} implies that two-component quantum Hall system can exhibit a fascinating intercomponent drag Hall conductance between pseudospin $\uparrow$ and pseudospin $\downarrow$, quantitatively determined by the symmetric matrix elements $C_{\uparrow\downarrow}=C_{\downarrow\uparrow}$. According to Laughlin's arguments, a charge Hall conductance allows the quantized charge pumping upon the adiabatic thread of one flux quantum~\cite{Laughlin1981}. To simulate this physical effect, we can utilize DMRG to calculate the charge pumping on infinite cylinder systems as the twisted angle $\theta_{\sigma}^{\alpha}$ changes~\cite{Gong2014}.

Numerically we cut the cylinder along the $x$ direction into two equal halves,
and the total charge of each component in the left part of the cylinder can be calculated from the expectation value of the particle number $N^L_{\sigma}(\theta_{\sigma'}^{y})=tr[\widehat{\rho}_L(\theta_{\sigma'}^{y})\widehat{N}^L_{\sigma}]$
(here $\widehat{\rho}_L$ the reduced density matrix of the left part, classified by the quantum numbers $\Delta Q_{\uparrow},\Delta Q_{\downarrow}$). Once the inserting flux quantum $\theta_{\sigma'}^{y}$ is adiabatically changed from zero to $2\pi$, we can obtain the evolution of the total charge of each component. The net charge transfer of the total charge of pseudospin $\sigma$ is now encoded by $N^L_{\sigma}(\theta_{\sigma'}^{y}=2\pi)-N^L_{\sigma}(\theta_{\sigma'}^{y}=0)$, manifesting the intercomponent drag charge Hall response, while $N^L_{\sigma'}(\theta_{\sigma'}^{y}=2\pi)-N^L_{\sigma'}(\theta_{\sigma'}^{y}=0)$ denotes the usual intracomponent Hall response. As illustrated in Fig.~\ref{pump}, for different cylinder widths under the flux thread $\theta_{\uparrow}=\theta,\theta_{\downarrow}=0,\theta\subseteq[0,2\pi]$, we find the charge pumpings
\begin{align}
  &\Delta N^L_{\uparrow}=|N^L_{\uparrow}(2\pi)-N^L_{\uparrow}(0)|\simeq C_{\uparrow\uparrow}=0, \nonumber\\
  &\Delta N^L_{\downarrow}=|N^L_{\downarrow}(2\pi)-N^L_{\downarrow}(0)|\simeq C_{\downarrow\uparrow}=1 \nonumber
\end{align}
in agreement with the calculation of the Chern number matrix in Eq.~\ref{chern} for two-component hardcore bosons at fillings $\nu=2$.

\subsection{Chiral edge modes}

\begin{figure}[t]
  \includegraphics[height=1.95in,width=3.3in]{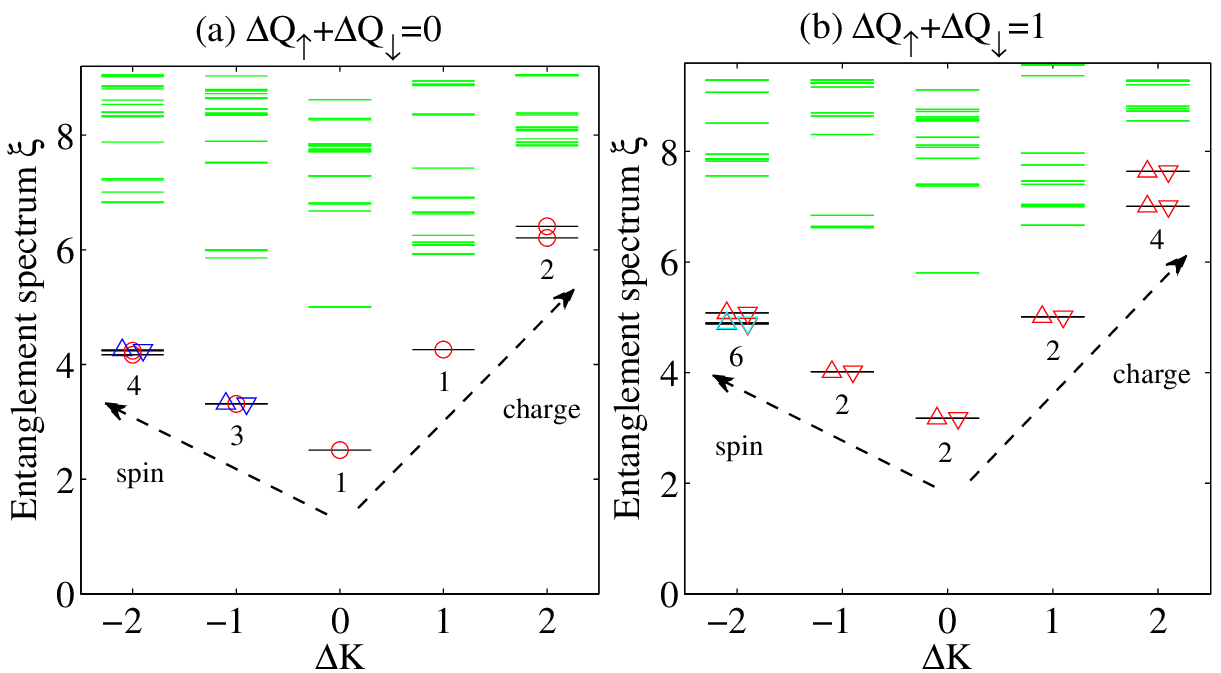}
  \caption{\label{es} (Color online) Chiral edge mode identified from the momentum-resolved entanglement spectrum for two-component hardcore bosons at $\nu_{\uparrow}=\nu_{\downarrow}=1$ with $U=\infty,V=0$ on the infinite $N_y=5$ cylinder of topological triangular lattice at two-third lattice filling in the typical total charge sectors (a) $\Delta Q_{\uparrow}+\Delta Q_{\downarrow}=0$ and (b) $\Delta Q_{\uparrow}+\Delta Q_{\downarrow}=1$. Here $\Delta Q_{\sigma}$ denotes the relative boson number $\Delta Q_{\sigma}=N^L_{\sigma}-N^{L_0}_{\sigma}$ of the left part of the infinite cylinder ($N^{L_0}_{\sigma}$ is the boson number of the state of $\widehat{\rho}_L$ with the largest eigenvalue). The horizontal axis shows the relative momentum $\Delta K=K_y-K_{y}^{0}$ (in units of $2\pi/N_y$). The numbers above the black dashed line label the level counting. In (a), the red circle denotes the level from the subsector $\Delta Q_{\uparrow}=\Delta Q_{\downarrow}=0$ and the blue upper (lower) triangle denotes the level from the subsector $\Delta Q_{\uparrow}-\Delta Q_{\downarrow}=\pm2$. In (b), the red upper (lower) triangle denotes the level from the subsector $\Delta Q_{\uparrow}-\Delta Q_{\downarrow}=\pm1$ and the green upper (lower) triangle denotes the level from the subsector $\Delta Q_{\uparrow}-\Delta Q_{\downarrow}=\pm3$.}
\end{figure}

For edge modes of Abelian FQH effect, they are described by free bosonic operators (namely collective density or current operators) which conserved the charge and spin quantum numbers. The characteristic chirality of edge modes and their level counting can be revealed through a low-lying entanglement spectrum in the bulk~\cite{Li2008,Wen1995}. By examining the structure of the momentum-resolved entanglement spectrum on the $N_y=5$ cylinder, we observe two counterpropagating branches of the low-lying bulk entanglement spectrum, consistent with the nonchiral nature of the BIQH phase.

In effective field theory, we can simulate the excitation level of the edge Hamiltonian $H_{edge}=2\pi/N_y\times v_s(L^c+L^s)$ in momentum space with the momentum shift operator $\Delta K=2\pi/N_y\times(L^c-L^s)$ for two counterpropagating modes. Here $L^{c(s)}=\sum_{j=1}^{\infty}jn_{j}^{c(s)}+(\Delta Q_{\uparrow}\pm\Delta Q_{\downarrow})^2/4$ and $n_{j}^{c(s)}$ denotes the set of non-negative integers describing harmonic oscillator modes. The pure charge or spin edge branch is determined by $(L^c\neq0,L^s=0)$ or $(L^c=0,L^s\neq0)$, and the value $(L^c,L^s)$ determines the energy level with a specific momentum $\Delta K$.
Therefore we can obtain the degeneracies of each edge mode (given by the total number of elements in the set $\{(n_{j}^{c},n_{j}^{s})\}$ with a given excitation energy) in any total charge sector $\Delta Q_{\uparrow}+\Delta Q_{\downarrow}$: for $\Delta Q_{\uparrow}+\Delta Q_{\downarrow}=0$, the level counting of the charge branch is $1,1,2,\cdots$ and the level counting of the spin branch is $1,3,4,\cdots$, as indicated in Fig.~\ref{es}(a); for $\Delta Q_{\uparrow}+\Delta Q_{\downarrow}=1$, the level counting of the charge branch is $2,2,4,\cdots$ and the level counting of the spin branch is $2,2,6,\cdots$, as indicated in Fig.~\ref{es}(b).

\subsection{Phase transition}

\begin{figure}[t]
  \includegraphics[height=1.9in,width=3.4in]{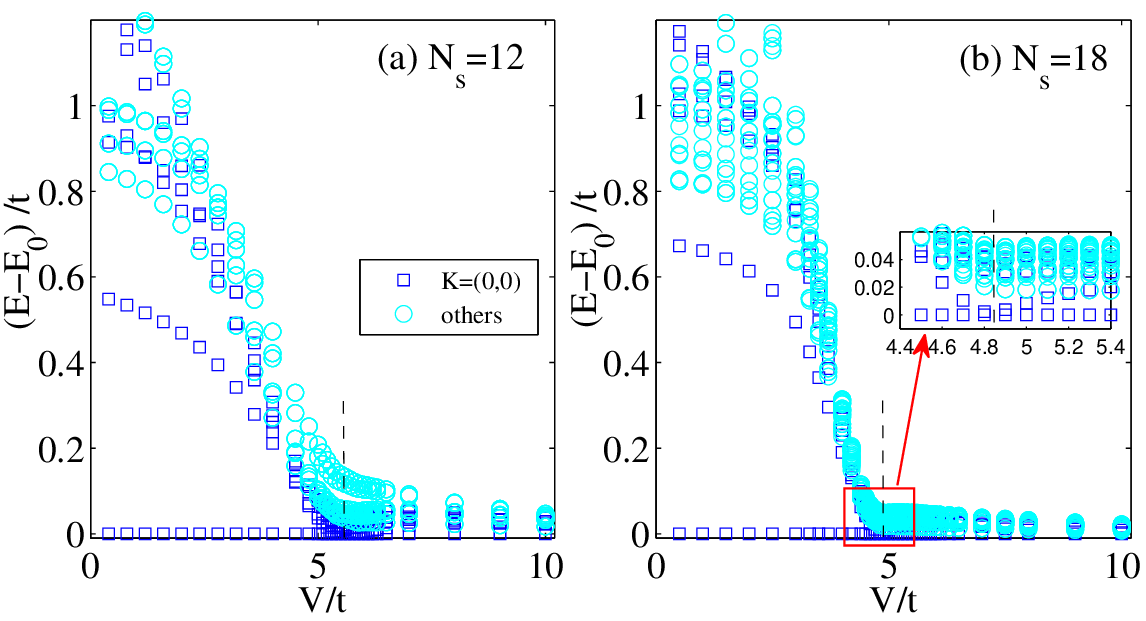}
  \caption{\label{interaction}(Color online) Numerical ED results for the low energy spectrum of two-component hardcore bosons at $\nu=2$ with infinite $U=\infty$ in topological triangular lattice as a function of nearest-neighboring interaction $V/t$. The system size (a) $N_s=3\times2\times2$ and (b) $N_s=3\times2\times3$. The black dashed line indicates the level-crossing point of the lowest ground state. The insets shows the zoom-in scan of the low energy spectrum near the transition point.}
\end{figure}

Finally, we turn to discuss the stability of the BIQH liquid when the nearest-neighboring repulsion $V$ is turned on. For single-component hardcore bosons in triangular lattice at commensurate lattice fillings $2/3$ or $1/3$, the $\sqrt{3}\times\sqrt{3}$ solid ordering pattern dominates~\cite{Wessel2005,Heidarian2005,Melko2005}. Here we also anticipate that for strong repulsion $V/t\gg1$, the BIQH liquid should be destroyed and the strongly interacting bosons would form a similar solid order pattern. Slightly different from the three-fold degeneracy of a solid order of single component bosons, for two-component bosons, the mutual exchange of the positions of intercomponent bosons leads to the same degenerate solid order due to the pseudospin-SU(2) symmetry of nearest-neighboring repulsion $V\sum_{\sigma,\sigma'}\!\sum_{\langle\rr,\rr'\rangle}n_{\rr',\sigma'}n_{\rr,\sigma}
=V\sum_{\langle\rr,\rr'\rangle}(n_{\rr',\uparrow}+n_{\rr',\downarrow})(n_{\rr,\uparrow}+n_{\rr,\downarrow})
=V\sum_{\langle\rr,\rr'\rangle}n_{\rr'}n_{\rr}$. In the ED study of the low energy spectrum as $V$ increases from zero as shown in Figs.~\ref{interaction}(a) and~\ref{interaction}(b) for different system sizes, we observe that the protecting gap of the BIQH ground state gradually collapses down to zero near a transition point $V=V_c$. We also verify that for small $V<V_c$, the density-density correlation $\langle n_{\rr'}n_{\rr}\rangle\approx2/3\times2/3$ is featureless and almost uniform as the distance $|\rr'-\rr|$ changes. However for large $V>V_c$, $\langle n_{\rr'}n_{\rr}\rangle$ displays the commensurate pattern $\langle n_{\rr'}n_{\rr}\rangle\simeq1$ when $n_{\rr'},n_{\rr}$ fall into the same sublattice, and the density-density structure factor $s(\qq)=\sum_{i,j}e^{i\qq\cdot(\rr'-\rr)}[\langle n_{\rr'}n_{\rr}\rangle-\langle n_{\rr'}\rangle\langle n_{\rr}\rangle]/N_s$ exhibits a strong peak at wave vector $\qq=0$, which indicates the solid order. Across the transition point, the ground state undergoes a level-crossing with the excited levels, and the system enters into the solid order phase for $V>V_c$ with a cluster of highly degenerate low-lying energy states.

\section{Conclusion}\label{summary}

To summarize, we have introduced a microscopic topological triangular lattice model of interacting two-component hardcore bosons as a possible realization of bosonic integer quantum Hall effect at filling fraction $\nu=2$. For strong onsite intercomponent Hubbard repulsion, we numerically demonstrate the emergence of the BIQH phase at two-third lattice filling in topological triangular lattice. Our comprehensive ED and DMRG simulations of different system sizes establish the inherent topological characteristics of this ground state: (i) the unique ground state degeneracy, (ii) quantized topological Chern number matrix $\mathbf{C}=\mathbf{K}^{-1}$, (iii) integer quantized intercomponent drag charge pumping without intracomponent Hall response, and (iv) two counterpropagating charge and spin edge branches, as predicted by  the $\mathbf{K}=\begin{pmatrix}
0 & 1\\
1 & 0\\
\end{pmatrix}$ matrix classification. Also we examine the robustness of the BIQH phase in the presence of nearest-neighboring repulsion, and claim the first-order phase transition between the BIQH phase and a solid order phase as nearest-neighboring repulsion increases from weak to strong. Along with the other efforts on explorations of BIQH effect in lattice models, our results contribute as a special example of the realization of the BIQH phase by tuning intercomponent repulsion.

\begin{acknowledgements}
T.S.Z thanks D. N. Sheng and W. Zhu for inspiring discussions and prior collaborations on BIQH effect in topological flat band models.
This work is supported by the National Natural Science Foundation of China (NSFC) under Grant No. 12074320.
\end{acknowledgements}

\end{document}